\documentclass[11pt,a4paper]{article}
\usepackage{amsmath}
\usepackage{ae}
\usepackage{graphicx}
\usepackage{multicol}
\usepackage{cancel}

\makeatletter

\newenvironment{figurehere}
{\def\@captype{figure}}
{}
\makeatother

\title{\raggedright Axion Electrodynamics in Topological Insulators for beginners. \\
\vspace{0.25cm}
\small
\noindent Part I: Introduction and basic equations \\
\noindent Part II: Images of a charge close to an interface ordinary insulator-topological insulator\\
\noindent Part III: Appendices on boundary conditions, the electrostatic image method and on electrodynamic atomic units}
\normalsize
\date{\today}
\author{Josep Planelles, Dept. Qu\'imica-F\'isica i Anal\'itica, Universitat Jaume I}
\begin{document}
\maketitle

\section{Introduction}
According to band theory, insulators are defined as those materials that have a gap between the occupied valence and the vacant conduction bands, while conductors do not have a  gap (either, because the last band is semi-occupied or because the conduction and valence bands overlap).\\

\noindent In 2006 unusual electromagnetic properties were observed in a  $CdTe|HgTe|CdTe$ quantum well:\cite{Bernevig} this quantum well behaves in bulk as an insulator but it was observed electric current across the interface, i.e., it behaves like a conductor in surface. The most striking feature is that this behavior cannot be accounted by the Maxwell's equations.\\

\noindent Earlier, in 1987, Frank Wilczek\cite{Wilczek87} suggested the possible behavior of this kind of materials and pointed out that it could be explained by means the axion electrodynamics that himself\cite{Wilczek78} and Steven Weinberg\cite{Weinberg} developed to understand the violation of combined symmetries of charge conjugation and parity in the strong interactions. The name axion is related to the name of the particle associated to this peculiar field.\\

\noindent In a relatively recent paper, Qi and Zhang\cite{Qi10} present and discuss experimental results on the $CdTe|HgTe|CdTe$ quantum well that shows an almost infinite resistance (i.e. behaves like an insulator) if the $HgTe$ thickness is smaller than a critical distance $d \sim 6.5 nm$, whereas its resistance  is small and displays the typical plateaus of the quantum Hall effect for $d > 6.5 nm$.\\

\noindent Axion electrodynamics can account for this behavior. To this end, a $\theta$ parameter, related to the Berry phase and the Chern number, is introduced. This parameter, called magneto-electric polarizability, is a piece-wise constant function. Its value is $\theta = 0$ for ordinary  and $\theta = \pi$ for time-reversal symmetry topological insulators, as e.g. $HgTe$. Axion electrodynamics with  $\theta(x, t) = 2 {\bf b \cdot x} - 2 b_0 t$ also describes Weyl semimetals\cite{Vazifeh} or, in general, the electrodynamics of magneto-electric media.\cite{ODell} A constant axion angle, $\theta(x, t) = k $, implies spatial and temporal translation symmetry conservation.\cite{Qi11}

\section{$\theta$-electrodynamics}
Electromagnetism in material media is described by Maxwell equations. In differential form and a.u. they can be written as (see a comment on electrodynamics atomic units in appendix 3):
$$
\begin{array}{lll}
{\boldsymbol \nabla}\cdot {\bf D}  = 4 \pi \rho \;{\rm \; \;\; (Gauss \, law)}&& {\boldsymbol \nabla} \times {\bf E} = -\frac{1}{c} \frac{\partial {\bf B}}{\partial t} {\rm \;\;\;\;\;\;\;\;\;(Faraday\, law)}
\\
\\
{\boldsymbol \nabla}\cdot  {\bf B} =0 \;\;\;\; \;\;\;\;{\rm (Gauss \, law \, for  \,magnetism)}&&{\bf \nabla} \times {\bf H} = \frac{1}{c}\frac{\partial{\bf D}}{\partial t}+\frac{4 \pi}{c} {\bf J}\;{\rm (Ampere \, law)}\\
\end{array}
$$
\noindent We should add to these equations the ${\bf D}, {\bf H}$ constitutive relations in terms of ${\bf E}, {\bf B}$. For linear material media these relations are simple: ${\bf D} = \epsilon {\bf E}$, ${\bf H} = \frac {{\bf B}} {\mu}$, with $\epsilon $, $\mu$ the dielectric constant and magnetic permeability. For isotropic materials $\epsilon$ and $\mu$ are constants, while in anisotropic materials are tensors, eventually coordinate-dependent. \\

\noindent In topological media, where magneto-electric effects take place, the electric displacement vector ${\bf D}$ is modified by the magnetic induction ${\bf B}$ and the magnetic field intensity ${\bf H}$ of is in turn influenced by the electric field. Then, the relations ${\bf D} = \epsilon {\bf E}$, ${\bf B} = \mu {\bf H}$ must be modified:\cite{MartinRuiz}

$$
{\bf D}=\epsilon  {\bf E} - \frac{\theta \alpha}{\pi} {\bf B} \;\;\;\;\;\;\;\;\; {\bf H}=\frac{{\bf B}}{\mu} + \frac{\theta \alpha}{\pi} {\bf E} 
$$
\noindent where $\alpha = 1/137$ is the fine-structure constant and $\theta$ an additional parameter that can be considered at the same level as the permittivity
 $\epsilon$ or the permeability $\mu$.  In topological media $\theta=\pi$ while in ordinary media $\theta=0$ (and we recover ordinary Maxwell equations).\\
 
\noindent These are the  ${\bf D}, {\bf H}$ constitutive equations as reported by Nogueira and van der Brink.\cite{Nogueira} It should be mentioned that opposite sign for the axion term can be found in the literature (see e.g. \cite{MartinRuizPRB}).  All the same, as pointed out by Vazifeh and Franz,\cite{Vazifeh} what allows the T- and P-invariant insulators to possess an axion term with $\theta=\pi$ is the $2\pi$ periodicity of the axion action in parameter $\theta$. Consequently, $\theta=\pi$ and $\theta=-\pi$ are two equivalent points and describe a T- and P-invariant system. Then, what about the axion term sign? As pointed out by Zirnstein and Rosenow\cite{Zirnstein} the response of a time-reversal-symmetric system always has to be time-reversal-symmetric. For this reason, the idea that the axion action describes the response of a finite time-reversal-symmetric topological insulator is a misconception and is incorrect. In other word, a topological system with periodic boundary conditions, the axion action generates no classical response, while in an open system i.e., a finite system separated by a border from ordinary material, its effect is canceled by the response of the topologically protected surface boundary state. Then, in order to get axion response we should break time-reversal. Actually, we need a setup such that time-reversal symmetry is broken only on the surface of the topological insulator, but is preserved in the bulk. This can be achieved e.g. by doping the topological insulator surface with magnetic impurities or by attaching on the surface a shell of another material with ordered magnetization (proximity effect). Then, since time-reversal symmetry is preserved in the bulk, we still get $\theta=\pi$. Meanwhile, since it is not on the surface, we can get axion response. In this case, the aforementioned axion sign is determined by the direction of the surface magnetization, $sign[\mathbf M \cdot \mathbf n]$, with $\mathbf n$  a surface unit vector pointing out of the topological insulator.\cite{Qi09,Sekine,Campos} \\

\noindent These modifications in the ${\bf D}, {\bf H}$ constitutive relations entail changes of two Maxwell equations, as we will show later. But first we will try to outline on the {\it natural} appearance of this extra term in electromagnetism. To this end, lets recall the electromagnetic energy expression:

$$
W=\frac{1}{2} ({\bf E}\cdot {\bf D}+{\bf B}\cdot {\bf H}) = \frac{1}{2} \epsilon E^2 +\frac{1}{2} \frac{1}{\mu} B^2
$$
\noindent On the other hand, the Lagrangian is:\cite{NotaLagrangiana}
$$
{\cal L}= \frac{1}{2} \epsilon E^2 - \frac{1}{2} \frac{1}{\mu} B^2
$$

\noindent From this Lagrangian, Maxwell equations can be obtained by means the Euler-Lagrange variational calculus (see e.g. Civelek et al.\cite{Civelek}).\\

\noindent We observe that the Lagrangian is quadratic with respect to the electromagnetic field. F. Wilczek\cite{Wilczek87} pointed out that there exists an additional quadratic term missing in the previous Lagrangian:

$$
\Delta {\cal L} = \kappa \, \theta \, {\bf E}\cdot {\bf B}
$$
\noindent By including this term, Wilczek obtains the axion electrodynamics equations. Here, however, we pursue a less elegant but simpler derivation: we incorporate, as already said, the ${\bf D}, {\bf H}$ constitutive relations in a $\theta \neq 0$ medium. By looking to Maxwell equations we can see:

\begin{align}
{\boldsymbol \nabla}\cdot {\bf D}  &= 4 \pi \rho  && {\rm  (changes \, as \, D \, changes)}  \label{eq1}\\
{\boldsymbol \nabla} \times {\bf E} &= -\frac{1}{c} \frac{\partial {\bf B}}{\partial t} && {\rm (does \, not  \, change)}  \label{eq2}\\
{\boldsymbol \nabla}\cdot  {\bf B} &=0 && {\rm (does \, not  \, change)}  \label{eq3}\\
{\bf \nabla} \times {\bf H} &= \frac{1}{c}\frac{\partial{\bf D}}{\partial t}+\frac{4 \pi}{c} {\bf J} && {\rm (changes \, as \, H\, and \, D\, changes)}  \label{eq4}
\end{align}

\noindent We add now the constitutive relations:\footnote{Incorporating the axion into Maxwell's
equations has the effect of "rotating" the electric and magnetic fields into each other:
$$
\left(\begin{matrix} {\bf E'}\\ c {\bf B'} \end{matrix}\right)=\left(\begin{matrix} {\bf E} - \frac{\theta \alpha}{\pi \epsilon} {\bf B}\\ c [{\bf B} +\frac{\theta \alpha}{\pi}\mu  {\bf E}]\end{matrix}\right)=\left(\begin{matrix} {\bf E} - [\frac{\theta \alpha}{\pi \epsilon c}]\; c {\bf B}\\ c {\bf B} +[\frac{\theta \alpha}{\pi}\mu c] \;{\bf E}\end{matrix}\right)=\frac{1}{\cos \xi} \left(\begin{matrix} \cos \xi & -\sin \xi \\ \sin \xi &\cos \xi\end{matrix}\right) \left(\begin{matrix} {\bf E}\\ c {\bf B} \end{matrix}\right)= \left(\begin{matrix} {\bf E}  -\tan \xi \; c {\bf B}\\ c {\bf B}+\tan \xi \; {\bf E} \end{matrix}\right)
$$
\noindent with $\tan \xi=\frac{\theta \alpha}{\pi \epsilon c} =\frac{\theta \alpha}{\pi}\frac{\sqrt{ \epsilon \mu} }{\epsilon}=\frac{\theta \alpha}{\pi}\sqrt{\frac{\mu}{\epsilon}}=\frac{\theta \alpha}{\pi}\mu \frac{1}{\sqrt{ \epsilon \mu}} =\frac{\theta \alpha}{\pi}\mu c $, the mixing angle $\xi$ depending then on the axion field strength $\theta$ and the coupling constants.}

\begin{align}
{\bf D}&=\epsilon  {\bf E} - \frac{\theta \alpha}{\pi} {\bf B} &&& \label{eq5}\\
{\bf H}&=\frac{{\bf B}}{\mu} + \frac{\theta \alpha}{\pi} {\bf E} &&& \label{eq6}
\end{align}

\noindent From equations (\ref{eq1}), (\ref{eq5}) we can write ${\boldsymbol \nabla}\cdot (\epsilon  {\bf E})=4 \pi \rho + \frac{\alpha}{\pi} \; {\boldsymbol \nabla}\cdot (\theta {\bf B})$. Since ${\boldsymbol \nabla}\cdot {\bf B}=0$, eq. (\ref{eq2}), we finally rewrite eq. (\ref{eq1}) as:\\
\begin{equation}
\label{eq1b}
{\boldsymbol \nabla}\cdot (\epsilon  {\bf E})=4 \pi \rho + \frac{ \alpha {\boldsymbol \nabla}\theta}{\pi} \cdot {\bf B}
\end{equation}

\noindent In a similar way, from equations (\ref{eq4}), (\ref{eq5}) and (\ref{eq6}) we have: \\

$${\bf \nabla} \times (\frac{1}{\mu} \, {\bf B})-\frac{1}{c}\frac{\partial{(\epsilon {\bf E})}}{\partial t}=\frac{4 \pi}{c} {\bf J}-\frac{1}{c}\frac{\alpha}{\pi} \;\frac{\partial (\theta {\bf B})}{\partial t}-\frac{\alpha}{\pi} \;{\boldsymbol \nabla} \times (\theta\,{\bf E})$$

\noindent Now we carry out the two last terms derivatives:\\
\begin{align}
-\frac{1}{c}\frac{\alpha}{\pi} \;\frac{\partial (\theta {\bf B})}{\partial t}-\frac{\alpha}{\pi} \;{\boldsymbol \nabla} \times (\theta {\bf E}) &=
-\frac{1}{c}\frac{\alpha}{\pi} \; \left(\frac{\partial \theta}{\partial t} \, {\bf B}+\theta \, \frac{\partial {\bf B}}{\partial t} \right)-
\frac{\alpha}{\pi} \; \left( {\boldsymbol \nabla} \theta \times{\bf E}+ \theta\; {\boldsymbol \nabla} \times{\bf E} \right) \nonumber\\
&= \frac{\alpha}{\pi} \,  \theta \, \left(-\frac{1}{c} \frac{\partial {\bf B}}{\partial t} -{\boldsymbol \nabla} \times {\bf E}\right) -
\frac{\alpha}{\pi} \, \left(\frac{1}{c} \frac{\partial \theta}{\partial t}\, {\bf B}+{\boldsymbol \nabla}\theta \times {\bf E} \right) \nonumber
\end{align}

\noindent Taking into account eq (\ref{eq2}),  the first bracket in the last equation must be zero. Then, eq (\ref{eq4}) turns into:

\begin{equation}
\label{eq4b}
{\bf \nabla} \times (\frac{1}{\mu} \, {\bf B})-\frac{1}{c}\frac{\partial{(\epsilon {\bf E})}}{\partial t}=\frac{4 \pi}{c} {\bf J}-\frac{1}{c}\frac{\alpha}{\pi}  \frac{\partial \theta}{\partial t}\, {\bf B} -\frac{\alpha}{\pi} \,{\boldsymbol \nabla}\theta \times {\bf E}
\end{equation}

\noindent The new equations (\ref{eq1b}) and (\ref{eq4b}), replacing eqs. (\ref{eq1}) and (\ref{eq4}), suggest the definition of effective charges and current densities. Thus, from (\ref{eq1b}) we define the effective charge $\rho_{\theta}$:

\begin{equation}
\label{eq9}
4 \pi \rho_{\theta} = \frac{ \alpha {\boldsymbol \nabla}\theta}{\pi} \cdot {\bf B} \to  \boxed{ \rho_{\theta} =\frac{\alpha}{4 \pi^2} \, {\boldsymbol \nabla}\theta \cdot {\bf B}}\\
\end{equation}

\noindent and from  (\ref{eq4b}) the effective current density ${\bf J}_{\theta}$:
\begin{equation}
\label{eq10}
\frac{4 \pi}{c} {\bf J}_{\theta} = \frac{\alpha}{\pi} \left(-\frac{1}{c} \frac{\partial \theta}{\partial t}\, {\bf B}- {\boldsymbol \nabla}\theta \times {\bf E}\right)
 \to  \boxed{ {\bf J}_{\theta} = \frac{\alpha}{4 \pi^2} \,  \left(-\frac{\partial \theta}{\partial t}\, {\bf B}-c\, {\boldsymbol \nabla}\theta \times {\bf E}\right) }\\
\end{equation}

\noindent Should $\theta=0$ then, equations (\ref{eq1b}) and (\ref{eq4b}) goes back to (\ref{eq1}) and (\ref{eq4}). Should be $\theta$ coordinates and time independent, i.e. should be $\theta$  a constant, then $\rho_{\theta}={\bf J}_{\theta}=0$ and we return again to the Maxwell equations.\\

\begin{center}
\begin{figurehere}
\resizebox{0.5\columnwidth}{!}{\includegraphics{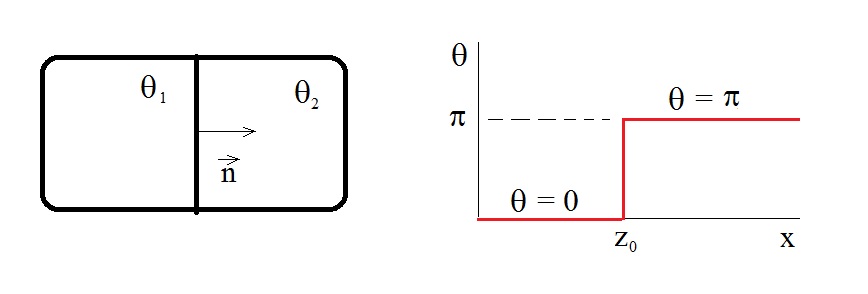}}
\end{figurehere}
\end{center}

\noindent As said above, $\theta=0$ in a vacuum or in an ordinary insulator, while $\theta=\pi$ for time reversal topological insulators. Therefore, $\theta$  is a piecewise constant function (see figure).\\

\noindent If $\theta(z)$ is a step function, then  ${\boldsymbol \nabla}\theta = \pi \delta(z_0) {\bf n}$, where ${\bf n}$ is a unit vector in the $z$ direction\footnote{Please note that both,  ${\boldsymbol \nabla}\theta$ and the unitary ${\bf n}$ vector, point from ordinary ($\theta =0$) to topological ($\theta =\pi$) insulator.} and $\delta(z_0)$ is the Dirac delta (see note\cite {nota1}). Therefore, since $\theta$ is time-independent, eq. (\ref{eq10}) leads to:

\begin{equation}
\label{eq11}
{\bf J}_{\theta} = -\frac{c\, \alpha}{4 \pi} \, \delta(z_0)\,  {\bf n} \times {\bf E}
\end{equation}

\noindent In a similar way, from eq. (\ref{eq9}) we have:

\begin{equation}
\label{eq12}
\rho_{\theta} = \frac{\alpha}{4 \pi} \, \delta(z_0)\,  {\bf n} \cdot {\bf B}
\end{equation}

\noindent The set of equations (\ref{eq1b}),  (\ref{eq2}),  (\ref{eq3}), (\ref{eq4b}) are those introduced by F. Wilczek\cite{Wilczek87} to define axion electrodynamics and are broadly used (see e.g. equation 39 in \cite{Sekine}, or equation 68 in \cite{Qi11} etc.). However, Luca Visinelli\cite{Visinelli} points out that Maxwell equations for an electromagnetic field show an internal symmetry, known as the duality transformation,
$$
\left( \begin{matrix} E' \\ B'\end{matrix}\right)=\left( \begin{matrix} \cos \xi & \sin \xi \\ -\sin \xi & \cos \xi \end{matrix}\right) \left( \begin{matrix} E \\ B\end{matrix}\right)
$$ 
\noindent and that whenever a pseudoscalar axion-like field  $\theta=\theta(x)$ is introduced in the theory, the dual symmetry is spontaneously and explicitly broken. He relates this broken symmetry to the fact that the introduction of an axion-like interaction with the electromagnetic field only modifies two of the four Maxwell equations (Gauss and Ampere laws), but not the remaining two equations (Faraday and Gauss law for B). The requirement that the electric and magnetic fields must satisfy also the above duality relation along with Gauss and Ampere laws for axion electrodynamics leads him we obtain new terms that also modify Faraday law and Gauss law for B (see eq. 26 in \cite{Visinelli}. See also \cite{nota2}).

%\begin{thebibliography}{99}

%\bibitem{Bernevig} B. Andrei Bernevig,  Taylor L. Hughes, and Shou-Cheng Zhang, Science 314 (2006) 1757. 
%\bibitem{Wilczek87} F. Wilczek, Phys. Rev. Lett. 58 (1987) 1799.
%\bibitem{Wilczek78} F. Wilczek, Phys. Rev. Lett. 40 (1978) 279.
%\bibitem{Weinberg} S. Weinberg, Phys. Rev. Lett. 40 (1978) 223.
%\bibitem{Qi10} X-L Qi and S-C Zhang, Physics Today 63 (2010) 33.
%\bibitem{Vazifeh} M. M. Vazifeh and M. Franz, Phys. Rev. Lett. 111 (2013) 027201.
%\bibitem{ODell} T. H. O'Dell, {\it The electrodynamics of magneto-electric media},  American Elsevier Pub. Co., New York 1970.
%\bibitem{Qi11} See e.g. section 3, after eq. 62, in  X-L Qi and S-C Zhang, Rev. Mod. Phys. 83 (2011) 1057.
%\bibitem{Civelek} C. Civelek  and T. F. Bechteler, Int. J. of Eng. Sci. 46 (2008) 1218.
%\bibitem{nota1} Dirac delta is defined as the derivative of the Heaviside function $\{ H(z>0)=1, \; H(z<0)=0\}$: $\delta(z)= \frac{d}{dz}H(z)$. Since $\theta= \pi\, H(z_0)$, then $\frac{d\theta}{dz}= \pi \, \frac{d H}{dz}= \pi \, \delta(z_0)$ and therefore  $ {\boldsymbol \nabla}\theta = \pi \delta(z_0) {\bf n}$.
%\bibitem{Sekine} A. Sekine and K. Nomura,  J. Appl. Phys. 129 (2021) 141101.
%\bibitem{Visinelli} L. Visinelli, Mod. Phys. Lett. A  28 (2013) 135062.
%\bibitem{nota2} Since for image charges calculation only equations (\ref{eq1b}) i (\ref{eq4b})  are employed, Visinelli proposal\cite{Visinelli} does not affect this kind of calculation.
%\end{thebibliography}

\newpage

\noindent {\Large \bf Part II: \\ 

\noindent Images of a charge close to an interface ordinary insulator-topological insulator}
\vspace{0.25cm}

\section{Image charge in topological insulators}
\subsection{Boundary conditions at the interface with a topological insulator}
In Appendix 1 we obtain the boundary conditions (BCs) in a.u. for electrostatics and magnetostatics (i.e., for time-independent fields). In particular, in absence of free charge and current, ${\bf D}_{2 \perp}={\bf D}_{1 \perp}$,  ${\bf B}_{2 \perp}={\bf B}_{1 \perp}$,  ${\bf H}_{2 \parallel}={\bf H}_{1 \parallel}$ and  ${\bf E}_{2 \parallel}={\bf E}_{1 \parallel}$. By injecting the constitutive relation (\ref{eq5}) in the first boundary condition we find:
\begin{equation}
\label{eq13a}
\epsilon_1 {\bf E}_{1 \perp}-\frac{\theta_1 \alpha}{\pi} {\bf B}_{1 \perp}=\epsilon_2 {\bf E}_{2 \perp}-\frac{\theta_2 \alpha}{\pi} {\bf B}_{2 \perp}
\end{equation}

\noindent Assuming the border surface at $z=0$, the second boundary condition, ${\bf B}_{2 \perp}(z=0)={\bf B}_{1 \perp}(z=0)=B_z$ yields,
\begin{equation}
\label{eq13b}
\epsilon_1 {\bf E}_{1 \perp}-\epsilon_2 {\bf E}_{2 \perp}= (\theta_1-\theta_2)\,\frac{\alpha}{\pi} B_z
\end{equation}
\noindent In particular,  if medium 1  ($z<0$) is ordinary ($\theta=0$) and medium 2 ($z>0$) topological ($\theta=\pi$), then

\begin{equation}
\label{eq13}
\epsilon_1 {\bf E}^{ord}_{1 \perp}=\epsilon_2 {\bf E}^{top}_{2 \perp}-\alpha B_z
\end{equation}

\noindent In a similar way, the third boundary conditions with the constitutive relation (\ref{eq6}) yield:

\begin{equation}
\label{eq14a}
\frac{1}{\mu_1} {\bf B}_{1 \parallel}+\frac{\theta_1 \alpha}{\pi} {\bf E}_{1 \parallel}=\frac{1}{\mu_2} {\bf B}_{2 \parallel}+\frac{\theta_2 \alpha}{\pi} {\bf E}_{2 \parallel}
\end{equation}

\noindent Assuming again the border surface at $z=0$, the fourth boundary condition, ${\bf E}_{2 \parallel}(z=0)={\bf E}_{1 \parallel}(z=0)={\bf E}_{\parallel}$, injected in eq. (\ref{eq14a}), yields,

\begin{equation}
\label{eq14b}
\frac{1}{\mu_1} {\bf B}_{1 \parallel}-\frac{1}{\mu_2} {\bf B}_{2 \parallel}=(\theta_2 -\theta_1)\frac{\alpha}{\pi} {\bf E}_{\parallel}
\end{equation}

\noindent In particular,  if medium 1  ($z<0$) is ordinary ($\theta=0$) and medium 2 ($z>0$) topological ($\theta=\pi$), then

\begin{equation}
\label{eq14}
\frac{1}{\mu_1} {\bf B}^{ord}_{1 \parallel}=\frac{1}{\mu_2} {\bf B}^{top}_{2 \parallel} + \alpha {\bf E}_{\parallel}.
\end{equation}

\subsection{Images of an electric charge close to the interface between ordinary and topological insulator}

We assume a static problem i.e., without magnetic induction ${\bf B}$ or electric field  ${\bf E}$ temporary dependence, and that there are no free charges and currents. Then, except at the interface, Maxwell equations (\ref{eq1}), (\ref{eq2}), (\ref{eq3}), (\ref{eq4}) become ${\boldsymbol \nabla} \cdot {\bf D} = 0$, ${\boldsymbol \nabla} \cdot {\bf B} = 0$, ${\boldsymbol \nabla} \times {\bf E} = 0$ and ${\boldsymbol \nabla} \times {\bf H} = 0$. Should the curl of a vector field be zero, then it can be written as the gradient of a scalar field. Therefore, from ${\boldsymbol \nabla} \times {\bf E} = 0$ we conclude that  ${\bf E}=-{\boldsymbol \nabla} V$. On the other hand, from  ${\boldsymbol \nabla} \times {\bf H} = 0$ and the constituent equation (\ref{eq6}), it follows that  ${\boldsymbol \nabla} \times \left( \frac{{\bf B}}{\mu} + \frac{\theta \alpha}{\pi} {\bf E} \right)= 0$ and then $\frac{{\bf B}}{\mu} + \frac{\theta \alpha}{\pi} {\bf E} =-{\boldsymbol \nabla} W$ or, in an equivalent way, ${\bf B}=-{\boldsymbol \nabla}\left[ \mu \, (W-\frac{\theta \alpha}{\pi} V)\right] = -{\boldsymbol \nabla} U$. Therefore, we can define {\it electric and magnetic potentials} whose opposite sign gradients yield the field. 

\begin{center}
\begin{figurehere}
\resizebox{0.22\columnwidth}{!}{\includegraphics{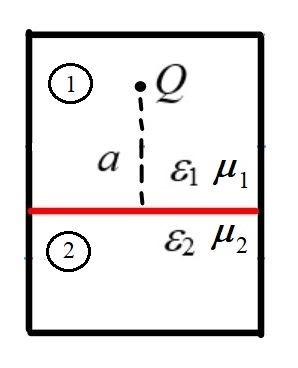}}
\end{figurehere}
\end{center}

\noindent Let $Q$ be an electric charge located at $(0,0,a)$ in an ($\epsilon_1, \mu_1$) ordinary medium in contact with an  ($\epsilon_2, \mu_2, \alpha$) topological insulator ( $\theta=0, \pi$ for ordinary and topological insulators, respectively). Assume the boundary at $z=0$. According to the image method (see Appendix 2), in order to calculate the potential in the ordinary insulator (zone 1, $z>0$) we must add a fictitious charge at $(0,0,-a)$ while to do it in the topological insulator (zone 2, $z<0$) we must add the fictitious charge at $(0,0,a)$. We have learned that a magnetic field ${\bf B}$ induces a surface charge $\sigma= \frac{\alpha}{4 \pi}\,  {\bf n} \cdot {\bf B}$ while an electric field (e.g. that from the electric source $Q$) generates surface currents ${\bf J}_s = -\frac{c\, \alpha}{4 \pi} \, {\bf n} \times {\bf E}$. The axial symmetry of the $Q$-generated electric field leads us to conclude from ${\bf J}_s$ formulae that circular currents are generated around the axis joining $Q$ with the interface (see Figure).

\begin{center}
\begin{figurehere}
\resizebox{0.55\columnwidth}{!}{\includegraphics{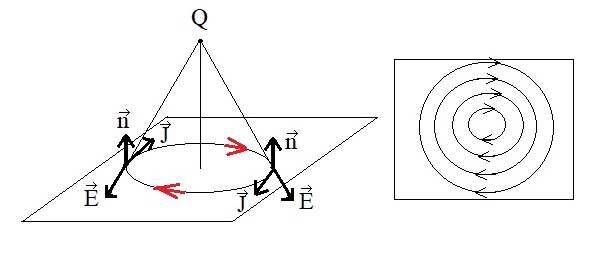}}
\end{figurehere}
\end{center}

\noindent This surface current, generated by $Q$ located at $z>0$, originates a magnetic field on the other side of the interface, $z<0$, proportional to the $Q$-electric field. Therefore proportional to $\frac{\bf r}{r^3}$. This magnetic field is equivalent to the magnetic field generated by a magnetic monopole $p_1$ located at $(0,0,b)$. Additionally, the surface current generates another magnetic field in the zone $z>0$,  also proportional to the electric field, equivalent to that generated by a magnetic monopole $p_2$ located, for sake of symmetry, at  $(0,0,-b)$.\\

\noindent The electric and magnetic potentials in both regions, according to the image method (Appendix 2) are:\footnote {The lack of symmetry of the magnetic field (axial vector) with respect to a horizontal plane lead us to write $p_1$ and $p_2$ as unknowns, with the expectation to find out $p_1=-p_2$.} 

\begin{align}
V(x,z>0) &= \frac{Q/\epsilon_1}{[x^2+(z-a)^2]^{1/2}}+ \frac{q}{[x^2+(z+a)^2]^{1/2}} \label{eqimag1}\\
V(x,z<0) &= \frac{Q/\epsilon_1}{[x^2+(z-a)^2]^{1/2}}+ \frac{q}{[x^2+(z-a)^2]^{1/2}} \label{eqimag2}\\
U(x,z>0) &= \frac{p_2}{[x^2+(z+b)^2]^{1/2}} \label{eqimag3}\\
U(x,z<0) &= \frac{p_1}{[x^2+(z-b)^2]^{1/2}} \label{eqimag4}
\end{align}

\noindent The ${\bf B}_{\perp}= -\frac{\partial U}{\partial z}$ continuity across the boundary means that $(\frac{\partial U(z>0)}{\partial z})_0 = (\frac{\partial U(z<0)}{\partial z})_0$. Then: 

\begin{align}
p_2\, \frac{(\cancelto{0}{z}+b)}{[x^2+(\cancelto{0}{z}+b)^2]^{3/2}} =p_1\, \frac{(\cancelto{0}{z}-b)}{[x^2+(\cancelto{0}{z}-b)^2]^{3/2}} \to  \boxed{ p_1=-p_2} \nonumber\\
\end{align}

\noindent Continuity at $z=0$ of ${\bf D}_{\perp}$, eq. (\ref{eq13}), i.e., $\epsilon_1 {\bf E}_{1 \perp}=\epsilon_2 {\bf E}_{2 \perp}-\alpha B_z$, with $B_z=-(\frac{\partial U}{\partial z})_0$ and $ {\bf E}_{\perp}=-(\frac{\partial V}{\partial z})_0$, leads to:\footnote{In the derivation we have calculated $B_z$ as the $U (z>0)$ derivative at $z=0$. The same result is achieved with $U (z<0)$ since a double change of sign (monopole charge and position) compensate.} 

\begin{eqnarray}
\epsilon_1 \, \left(\frac{Q}{\epsilon_1} \frac{(\cancelto{0}{z}-a)}{[x^2+(\cancelto{0}{z}-a)^2]^{3/2}} +q \, \frac{(\cancelto{0}{z}+a)}{[x^2+(\cancelto{0}{z}+a)^2]^{3/2}}\right) &=& \epsilon_2 \, \left(\frac{Q}{\epsilon_1} \frac{(\cancelto{0}{z}-a)}{[x^2+(\cancelto{0}{z}-a)^2]^{3/2}} +q \, \frac{(\cancelto{0}{z}-a)}{[x^2+(\cancelto{0}{z}-a)^2]^{3/2}}\right) \nonumber\\
&&- \alpha \, p_1 \,  \frac{(\cancelto{0}{z}+b)}{[x^2+(\cancelto{0}{z}+b)^2]^{3/2}} \nonumber\\
\nonumber\\
\to Q\, (1- \frac{\epsilon_2}{\epsilon_1}) \frac{(-a)}{[x^2+a^2]^{3/2}}+\frac{a \, q}{[x^2+a^2]^{3/2}}\, (\epsilon_1+\epsilon_2)&=&-\frac{b}{[x^2+b^2]^{3/2}}\, \alpha\, p_1 \;\;\forall x \nonumber\\
\nonumber\\
&\to& \boxed{a=b}\nonumber\\
&&\nonumber\\
\to  Q\, ( \frac{\epsilon_2-\epsilon_1}{\epsilon_1})+ (\epsilon_1+\epsilon_2) \, q &=& -\alpha\, p_1 \to \boxed{p_1=-\frac{Q}{\alpha} \frac{\epsilon_2-\epsilon_1}{\epsilon_1}-\frac{q}{\alpha} \; (\epsilon_1+\epsilon_2)}\label{eq15}
\end{eqnarray}

\noindent The ${\bf H}_{\parallel}$ continuity at $z=0$, eq. (\ref{eq14}), i.e., $\frac{1}{\mu_1} {\bf B}_{1 \parallel}=\frac{1}{\mu_2} {\bf B}_{2 \parallel} + \alpha {\bf E}_{\parallel}$, with $p_1=-p_2$, ${\bf B}_{\parallel}=-(\frac{\partial U}{\partial x})_{z=0}$, ${\bf E}_{\parallel}=-(\frac{\partial V}{\partial x})_{z=0}$, and  $(\frac{\partial V(z>0)}{\partial x})_{z=0} = (\frac{\partial V(z<0)}{\partial x})_{z=0}$, yields (at $z=0$):

\begin{eqnarray}
\frac{1}{\mu_1} \frac{x \, p_1}{[x^2+a^2]^{3/2}} &=&\frac{1}{\mu_2} \frac{x \, (-p_1)}{[x^2+a^2]^{3/2}}+ \alpha \;\left( \frac{Q}{\epsilon_1} \frac{x}{[x^2+a^2]^{3/2}}+ q \;\frac{x}{[x^2+a^2]^{3/2}} \right)\nonumber\\
\nonumber\\
\to (\frac{1}{\mu_1}+\frac{1}{\mu_2}) p_1&=&\alpha \; (\frac{Q}{\epsilon_1}+q)\nonumber\\
\nonumber\\
&\to& \boxed{p_1= \frac{\alpha}{\frac{1}{\mu_1}+\frac{1}{\mu_2}} \;(\frac{Q}{\epsilon_1}+q)} \label{eq16}
\end{eqnarray}

\noindent From equations (\ref{eq15}) and (\ref{eq16}) it follows:
\begin{equation}
\begin{array}{l}
\frac{Q}{\alpha} \frac{\epsilon_2-\epsilon_1}{\epsilon_1}+\frac{q}{\alpha} (\epsilon_1+\epsilon_2) = - \frac{\alpha}{\frac{1}{\mu_1}+\frac{1}{\mu_2}} \;(\frac{Q}{\epsilon_1}+q)\\
\\
\to \frac{Q}{\epsilon_1}\left( \frac{\epsilon_2-\epsilon_1}{\alpha} +\frac{\alpha}{\frac{1}{\mu_1}+\frac{1}{\mu_2}}\right)+ q\; \left( \frac{\epsilon_1+\epsilon_2}{\alpha} +\frac{\alpha}{\frac{1}{\mu_1}+\frac{1}{\mu_2}}\right)=0\\
\\
\to q = -  \frac{Q}{\epsilon_1} \frac{\frac{\epsilon_2-\epsilon_1}{\alpha}+\frac{\alpha}{\frac{1}{\mu_1}+\frac{1}{\mu_2}}}{\frac{\epsilon_2+\epsilon_1}{\alpha}+\frac{\alpha}{\frac{1}{\mu_1}+\frac{1}{\mu_2}}} \nonumber\\
\end{array}
\end{equation}
\begin{equation}
\begin{array}{l}
\to  \boxed{q = \frac{Q}{\epsilon_1} \frac{(\frac{1}{\mu_1}+\frac{1}{\mu_2}) (\epsilon_1-\epsilon_2)-\alpha^2}{(\frac{1}{\mu_1}+\frac{1}{\mu_2}) (\epsilon_1+\epsilon_2)+\alpha^2}}   \label{eq17}
\end{array}
\end{equation}

\noindent By replacing $q$, eq. (\ref{eq17}), in eq. (\ref{eq16}) and calling $M=\frac{1}{\mu_1}+\frac{1}{\mu_2}$, we find:
\begin{equation}
\begin{array}{l}
p_1= \frac{\alpha}{M} \;(\frac{Q}{\epsilon_1}+\frac{Q}{\epsilon_1} \frac{M \, (\epsilon_1-\epsilon_2)-\alpha^2}{M \, (\epsilon_1+\epsilon_2)-\alpha^2})=
 \frac{\alpha}{M} \;\frac{Q}{\epsilon_1} (1+ \frac{ (\epsilon_1-\epsilon_2)-\alpha^2/M}{(\epsilon_1+\epsilon_2)+\alpha^2/M})\\
\\
= \frac{\alpha}{M} \;\frac{Q}{\epsilon_1} \frac{(\epsilon_1+\epsilon_2)+\alpha^2/M+(\epsilon_1-\epsilon_2)-\alpha^2/M}{(\epsilon_1+\epsilon_2)+\alpha^2/M}=
 \frac{\alpha Q}{\epsilon_1} \frac{2 \epsilon_1}{M (\epsilon_1+\epsilon_2)+\alpha^2} \nonumber
\end{array}
\end{equation}
\begin{equation}
\begin{array}{l}
\to \boxed{p_1= \alpha \frac{2 Q}{(\frac{1}{\mu_1}+\frac{1}{\mu_2})(\epsilon_1+\epsilon_2)+\alpha^2}} \label{eq18}
\end{array}
\end{equation}

\noindent Eqs. (\ref{eq17}) i (\ref{eq18}) provide the value of the $Q$-induced electric charge and magnetic monopole originated as images across the ordinary-topological insulator interface.\footnote {By identifying $\alpha^2$ with  $4 P_3^2 \alpha^2$ (for $P_3=\pm \frac{1}{2}$ i.e.  $4 P_3^2=1$, see \cite{Qi09} page 1185) we can check that Eq. (\ref{eq17}) matches eq. 3.2 and eq. (\ref{eq18}) matches eq. 3b in \cite{Qi09}.} 

\subsection{Images of a magnetic charge close to the interface between ordinary and topological insulator}
Beyond the existence or not of a magnetic monopole, it is worth getting its electric and magnetic images across an ordinary-topological interface. On the one hand because two monopole at a given distance constitute a dipole, and doubtless magnetic dipoles do exist, so that monopole image and the  superposition principle can be used to calculate dipole images. Additionally, an electric charge between two interfaces yields a cascade of image charges,\cite{Kumagai} so it can be useful in this and other kind of more elaborate calculations.\\

\noindent We have already pointed out that the electric field of a charge in the vicinity of an interface with a topological insulator generates a surface current, eq. (\ref{eq11}), ${\bf J}_{\theta} = -\frac{c \alpha}{4 \pi} \delta(z_0) {\bf n} \times {\bf E}$ (it is implied that the interface is located at $z_0$). This current is the same as that generated by a magnetic monopole, symmetrically located with respect to the source electric charge $Q$, having a magnetic charge, eq. (\ref{eq18}), $p=   \frac{2 \alpha}{(\frac{1}{\mu_1}+\frac{1}{\mu_2})(\epsilon_1+\epsilon_2)+\alpha^2} \, Q$.\\  

\noindent Similarly, a magnetic monopole in the vicinity of a $z_0$-located interface with a topological insulator creates a magnetic field $\bf B$  which in turn generates a surface charge density, eq. (\ref{eq12}), $\rho_{\theta} = \frac{\alpha}{4 \pi} \, \delta(z_0)\,  {\bf n} \cdot {\bf B}$.\\ 

\noindent As the monopole generates a radial field (like that of a charge or electric monopole, see Figure), the charge density induced by the magnetic monopole will generate an electric potential like that of an electric charge $q$ located at the position determined the image method.

\begin{center}
\begin{figurehere}
\resizebox{0.35\columnwidth}{!}{\includegraphics{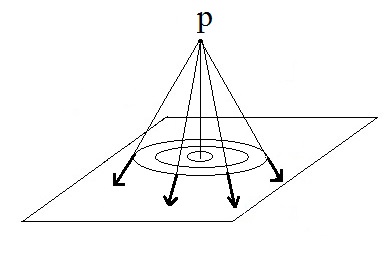}}
\end{figurehere}
\end{center}

\noindent Like images of an electric charge, a magnetic monopole $P$ generates an electric ($V$) and magnetic ($U$) potentials given by:\\

\begin{align}
U_1(x,z>0) &= \frac{\mu_1 \, P}{[x^2+(z-a)^2]^{1/2}}+\frac{p_2}{[x^2+(z+a)^2]^{1/2}} \label{mag1}\\
U_2(x,z<0) &= \frac{\mu_1 \, P}{[x^2+(z-a)^2]^{1/2}} +\frac{p_1}{[x^2+(z-a)^2]^{1/2}} \label{mag2}\\
V_1(x,z>0) &= \frac{q}{[x^2+(z+b)^2]^{1/2}} \label{mag3}\\
V_2(x,z<0) &=  \frac{q}{[x^2+(z-b)^2]^{1/2}} \label{mag4}\\
\end{align}

\noindent with boundary conditions (BCs):

\begin{align}
\epsilon_1 \left( \frac{\partial V_1}{\partial z}\right)_0 &= \epsilon_2 \left( \frac{\partial V_2}{\partial z}\right)_0- \alpha \,  \left( \frac{\partial U}{\partial z}\right)_0 \label{mag5}\\
\frac{1}{\mu_1} \left( \frac{\partial U_1}{\partial x}\right)_0 &= \frac{1}{\mu_2} \left( \frac{\partial U_2}{\partial x}\right)_0 + \alpha \,  \left( \frac{\partial V}{\partial x}\right)_0 \label{mag6}
\end{align}
\noindent where since $\left( \frac{\partial U_1}{\partial z}\right)_0 = \left( \frac{\partial U_2}{\partial z}\right)_0$ and $ \left( \frac{\partial V_1}{\partial x}\right)_0 =\left( \frac{\partial V_2}{\partial x}\right)_0$, we can employ  $U$ and $V$ of either region.\\

\noindent From the first BC, eq. (\ref{mag5}), we have:
\begin{equation}
\label{mag7}
\frac{\epsilon_1 \, q \, b}{[x^2+b^2]^{3/2}}=\frac{-\epsilon_2 \, q \, b}{[x^2+b^2]^{3/2}}+\alpha \, \left( \frac{\mu_1 \, P \, a}{[x^2+a^2]^{3/2}}+\frac{p_1 \, a}{[x^2+a^2]^{3/2}}\right) \; \forall x \; \to \boxed{a=b}
\end{equation}
\noindent and coming back to this equation with $a=b$ it follows:
\begin{equation}
\label{mag8}
\epsilon_1 \, q =-\epsilon_2 \, q + \, \alpha \, \left(\mu_1 \, P + p_1 \right) \;  \to \boxed{ q = \frac{\alpha}{\epsilon_1+\epsilon_2} \,(p_1+\mu_1 \, P) }
\end{equation}

\noindent From the second BC, eq. (\ref{mag6}), we find out:

\begin{align}
\label{mag9}
\frac{1}{\mu_1} \left( \frac{\mu_1 \,P \,x}{[x^2+a^2]^{3/2}}+\frac{p_2 \,x}{[x^2+a^2]^{3/2}}\right) = \frac{1}{\mu_2} \left( \frac{\mu_1 \,P \,x}{[x^2+a^2]^{3/2}}+\frac{p_1 \,x}{[x^2+a^2]^{3/2}}\right)+ \alpha \, \frac{q \, x}{[x^2+a^2]^{3/2}} \;\;\;\; \forall x  \nonumber\\
\nonumber\\
\to P+\frac{p_2}{\mu_1}=\frac{\mu_1}{\mu_2}\, P + \frac{1}{\mu_2} \, p_1 + \alpha \, q \; \to \boxed{P\; \left(1-\frac{\mu_1}{\mu_2}\right)+\frac{p_2}{\mu_1}-\frac{p_1}{\mu_2}-\alpha \; q = 0}
\end{align}

\noindent The boundary condition ${\bf E}_{2 \parallel}={\bf E}_{1 \parallel}$ turns to be an identity which brings nothing. Finally the boundary condition ${\bf B}_{1 \perp} = {\bf B}_{2 \perp}$ with  ${\bf B}_{\perp} =-\frac{\partial U}{\partial z}$ leads to:

\begin{align}
\label{mag10}
\frac{\mu_1 \, P \, (\cancelto{0}{z}-a)}{[x^2+a^2]^{3/2}}+\frac{p_2 \, (\cancelto{0}{z}+a)}{[x^2+a^2]^{3/2}} = \frac{\mu_1 \, P \, (\cancelto{0}{z}-a)}{[x^2+a^2]^{3/2}} + \frac{p_1 \, (\cancelto{0}{z}-a)}{[x^2+a^2]^{3/2}} & \to \boxed{p_1=-p_2}
\end{align}

\noindent Now, by combining eqs. (\ref{mag9}) and  (\ref{mag10}), taking into account (\ref{mag8}), we find:
\begin{align}
\label{mag11}
P \, \left(1-\frac{\mu_1}{\mu_2}\right)-p_1 \left(\frac{1}{\mu_1}+\frac{1}{\mu_2}\right)-\frac{\alpha^2}{\epsilon_1+\epsilon_2}\, \left( p_1+\mu_1 \, P\right)=0 \nonumber\\
\nonumber\\
\to P \, \left(1-\frac{\mu_1}{\mu_2}-\frac{\alpha^2\, \mu_1}{\epsilon_1+\epsilon_2}\right)-p_1\left(\frac{1}{\mu_1}+\frac{1}{\mu_2}+\frac{\alpha^2}{\epsilon_1+\epsilon_2} \right)=0\nonumber\\
\nonumber\\
\to p_1= P \;\frac{1-\frac{\mu_1}{\mu_2}-\mu_1 \,\frac{\alpha^2}{\epsilon_1+\epsilon_2}}{\frac{1}{\mu_1}+\frac{1}{\mu_2}+\frac{\alpha^2}{\epsilon_1+\epsilon_2}}
\to \boxed{ p_1=  \mu_1 \, P \; \frac{\frac{1}{\mu_1}-\frac{1}{\mu_2}-\frac{\alpha^2}{\epsilon_1+\epsilon_2}}{\frac{1}{\mu_1}+\frac{1}{\mu_2}+\frac{\alpha^2}{\epsilon_1+\epsilon_2}} }
\end{align}
\noindent Note that if the insulator is not topological but ordinary, then we should take $\alpha=0$ and obtain $ p_1=  \mu_1 \, P \; \frac{\mu_2-\mu_1}{\mu_1+\mu_2}$ which is the image of a monopole in front of an interface between two different ordinary media.\\

\noindent By injecting $p_1$, eq. (\ref{mag11}), into $q$ given by equation (\ref{mag8}) we get:

\begin{align}
\label{mag12}
q = \frac{\alpha}{\epsilon_1+\epsilon_2} \,\mu_1 \, P \ \left(\frac{\frac{1}{\mu_1}-\frac{1}{\mu_2}-\frac{\alpha^2}{\epsilon_1+\epsilon_2}}{\frac{1}{\mu_2}+\frac{1}{\mu_1}+\frac{\alpha^2}{\epsilon_1+\epsilon_2}} + 1\right) 
\end{align}

\noindent We can simplify eqs. (\ref{mag11}) i (\ref{mag12}) as follow:
\begin{align}
\label{mag13}
\boxed {p_1 =  \mu_1 \, P \; \frac{(\frac{1}{\mu_1}-\frac{1}{\mu_2}) (\epsilon_1+\epsilon_2)-\alpha^2}{(\frac{1}{\mu_1}+\frac{1}{\mu_2}) (\epsilon_1+\epsilon_2)+\alpha^2}}
\end{align}
\begin{align}
\label{mag14}
\boxed {q = \alpha \; \frac{2 \, P}{(\frac{1}{\mu_1}+\frac{1}{\mu_2}) (\epsilon_1+\epsilon_2)+\alpha^2}}
\end{align}

\noindent The result is a perfect analogy of the image of an electric charge, with the detail that the image $q$ of the charge $Q$, eq. (\ref{eq17}), is proportional to $ (\frac{1}{\mu_1}+\frac{1}{\mu_2}) (\epsilon_1-\epsilon_2)$ while the image $p$ of the monopole $P$, eq. (\ref{mag14}), is proportional to $ (\frac{1}{\mu_1}-\frac{1}{\mu_2}) (\epsilon_1+\epsilon_2)$. Dielectric constants and magnetic permeabilities exchange their role.\\

\noindent {\bf Exercise}\\
\noindent An electric charge is located in the central region of a quantum well build up by a topological insulator surrounded by an ordinary insulator. (a) Calculate the electrical potential generated in the central region, assuming that both the insulator of the central region and that of the surrounding barriers, are ordinary. Check that the result agrees with equations (2.8) and (2.14) by Kumagai and Takagahara.\cite{Kumagai} (b) Calculate the electrical and magnetic potentials under the assumption that the insulator of the central region is topological. Check that by forcing  $\alpha=0$ the magnetic potential goes to zero and the electric potential matches that obtained in the previous section.

\newpage
\noindent {\Large \bf Part III}: \\
\section{Appendix 1: Boundary conditions at the interface of ordinary insulators}
Maxwell's equations in integral form (and MKS rational system) are:
\begin{equation}
\begin{array}{lll}
\iint {\bf D} \cdot d {\bf S} =\iiint\rho \;dv && \iint {\bf \nabla} \times {\bf E} \cdot d {\bf S}= \oint {\bf E}\cdot d {\boldsymbol \ell}=-\iint\frac{\partial {\bf B}}{\partial t} \cdot d {\bf S}
\\
\iint {\bf B} \cdot d{\bf S} =\iiint {\bf \nabla} \cdot {\bf B} \;dv =0 && \iint {\bf \nabla} \times {\bf H} \; d{\bf S}= \oint {\bf H}\cdot d{\boldsymbol \ell}=\iint {\bf J} \cdot d{\bf S}+\iint\frac{\partial{\bf D}}{\partial t} \cdot d{\bf S}\\
\end{array}
\end{equation}
\noindent Should we consider time independence, the last two equations  including time derivatives are simplified. Then, Maxwell equations become:
\begin{equation}
\begin{array}{lll}
\iint {\bf D} \cdot d {\bf S} =\iiint\rho \;dv &&  \oint {\bf E}\cdot d {\boldsymbol \ell}=0
\\
\iint {\bf B} \cdot d{\bf S} =0 && \oint {\bf H}\cdot d{\boldsymbol \ell}=\iint {\bf J} \cdot d{\bf S}\\
\end{array}
\end{equation}

\begin{center}
\begin{figurehere}
\resizebox{0.4\columnwidth}{!}{\includegraphics{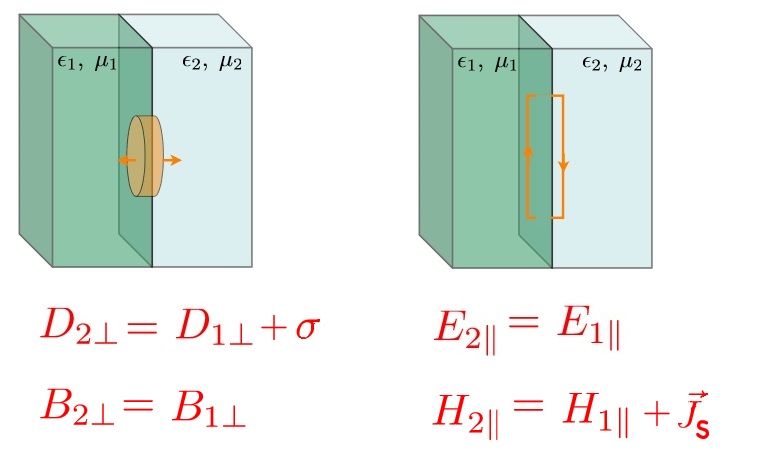}}
\end{figurehere}
\end{center}

\noindent On left in the Figure we show a slab of area $\Delta S$ and infinitesimal height $dx$. We rewrite the integral $\iint {\bf B} \cdot d{\bf S}=0$ for this slab as:

$$
0=\iint {\bf B} \cdot d{\bf S}= \iint {\bf B} \cdot {\bf n}\, dS = ({\bf B}_2\cdot {\bf n}-{\bf B}_1\cdot {\bf n})\, \Delta S=({\bf B}_2-{\bf B}_1)\cdot {\bf n}\, \Delta S \to \boxed{{\bf B}_{2 \perp}={\bf B}_{1 \perp}}
$$
\noindent In a similar way, $\iint {\bf D} \cdot d{\bf S}=({\bf D}_2-{\bf D}_1)\cdot {\bf n} \,\Delta S$. On the other hand $\iiint\rho \;dv =\iiint\rho \, dx \, \Delta S= \sigma \Delta S$. Then, 
$$({\bf D}_2-{\bf D}_1)\cdot {\bf n} =\sigma  \to   \boxed{ {\bf D}_{2 \perp}={\bf D}_{1 \perp} +\sigma } $$

\noindent On the right hand side of the figure it is drawn a circuit in the shape of a rectangle with a long side $\Delta {\boldsymbol \ell}$  and a narrow infinitesimal side $dx$. The circulation of the electric field $\oint {\bf E}\cdot d {\boldsymbol \ell}=0$ in this circuit is:

$$
0=\oint {\bf E}\cdot d {\boldsymbol \ell}=({\bf E}_2-{\bf E}_1)\cdot \Delta {\boldsymbol \ell} \to   \boxed{{\bf E}_{2 \parallel}={\bf E}_{1 \parallel}}
$$
\noindent In a similar way, $\oint {\bf H}\cdot d{\boldsymbol \ell}=({\bf H}_2-{\bf H}_1)\cdot \Delta \ell$. Also, 
$\iint {\bf J} \cdot d{\bf S}=\int {\bf J} \cdot (dx\,\Delta \ell) {\boldsymbol \tau}=(\int {\bf J} dx)\cdot {\boldsymbol \tau} \Delta \ell  = {\bf J}_s  \Delta  \ell$, where $J_s$ represents surface current at the interface. Therefore:
$$
 \boxed{ {\bf H}_{2 \parallel}={\bf H}_{1 \parallel} +{\bf J}_s  }
$$

\subsection{Boundary conditions at the interface of ordinary insulators in a.u.}
Maxwell equations in differential form and a.u. reads: 
$$
\begin{array}{lll}
{\boldsymbol \nabla}\cdot {\bf D}  = 4 \pi \rho \;{\rm \; \;\; (Gauss \, law)}&& {\boldsymbol \nabla} \times {\bf E} = -\frac{1}{c} \frac{\partial {\bf B}}{\partial t} {\rm \;\;\;\;\;\;\;\;\;(Faraday\, law)}
\\
\\
{\boldsymbol \nabla}\cdot  {\bf B} =0 \;\;\;\; \;\;\;\;{\rm (Gauss \, law \, for  \,magnetism)}&&{\bf \nabla} \times {\bf H} = \frac{1}{c}\frac{\partial{\bf D}}{\partial t}+\frac{4 \pi}{c} {\bf J}\;{\rm (Ampere \, law)}\\
\end{array}
$$

\noindent Therefore,  $4 \pi$ must be added to the ${\bf D}$ boundary condition and $\frac{4\pi}{c}$ to the ${\bf H}$ boundary condition: 
$$
 \boxed{ {\bf D}_{2 \perp}={\bf D}_{1 \perp} +4 \pi \sigma } \;\;\;\; \;\;\;\;  \boxed{ {\bf H}_{2 \parallel}={\bf H}_{1 \parallel} +\frac{4\pi}{c} \,{\bf J}_s  }
$$

\section{Appendix 2: Electrostatic Image Method}
Let $Q$ be a charge in a dielectric constant $\epsilon_1$ medium. $Q$ polarizes this medium. Should the medium be infinite then the result is that $Q$ plus the medium generate the same electric field as that of an effective charge $Q/\epsilon_1$  in a vacuum. If there is an interface separating two different polarizability media, i.e., with different dielectric constant (see figure), on the surface of the upper medium appears the lower ends of the last dipole layer induced in this medium which are not compensated with the first dipole layer induced in the lower medium. The result is that the charge $Q$  located at a distance $a$ from the interface plus the two media generate the same field as that generated in a vacuum by an effective charge $Q/\epsilon_1$ located where the charge $Q$ is located plus a surface density $\sigma$ located at the interface between the media. The charge distribution $\sigma$ generates in the upper medium the same field as that generated by a charge $q_1$ located symmetrically in the lower medium at a distance $b$, while in the lower medium it generates the same field than that  generated by a charge $q_2$ located in the same vertical as $Q$ (for symmetry reasons) at a distance $c$ from the interface.

\begin{center}
\begin{figurehere}
\resizebox{0.9\columnwidth}{!}{\includegraphics{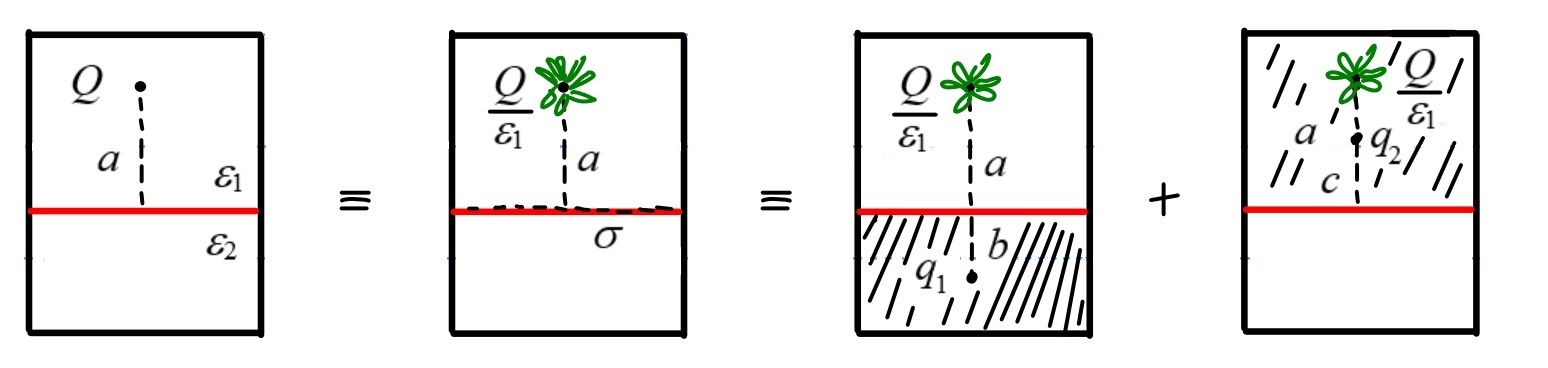}}
\end{figurehere}
\end{center}

\noindent The electric potential in the upper medium $V(z>0)$ is generated by $Q/\epsilon_1$ located at $z_Q=a$ plus $q_1$ located at $z_{q_1}=-b$:
\begin{equation}
V(z>0) = \frac{Q/\epsilon_1}{[x^2+(z-a)^2]^{1/2}}+\frac{q_1}{[x^2+(z+b)^2]^{1/2}}
\end{equation} 
\noindent The electric potential in the lower medium $V(z<0)$ is generated by $Q/\epsilon_1$ located at $z_Q=a$ and  $q_2$ located a $z_{q_2}=c$:
\begin{equation}
V(z<0) = \frac{Q/\epsilon_1}{[x^2+(z-a)^2]^{1/2}}+\frac{q_2}{[x^2+(z-c)^2]^{1/2}}
\end{equation}
\noindent To determine the values $b,c,q_1,q_2$ we apply the BCs obtained in Appendix 1. From BC ${\bf E}_{2 \parallel}={\bf E}_{1 \parallel}$, with ${\bf E}_{\parallel}=-\frac{\partial V}{\partial x}$ we find:

\begin{equation}
{\bf E}_{1 \parallel}(z>0) = \frac{Q}{\epsilon_1} \frac{x}{[x^2+(z-a)^2]^{3/2}}+q_1 \; \frac{x}{[x^2+(z+b)^2]^{3/2}}
\end{equation}
\begin{equation}
{\bf E}_{2 \parallel}(z<0) = \frac{Q}{\epsilon_1} \frac{x}{[x^2+(z-a)^2]^{3/2}}+q_2 \; \frac{x}{[x^2+(z-c)^2]^{3/2}}
\end{equation}

\noindent Equating the two equations at $z=0$ we find that: 
\begin{equation}
\frac{q_1}{[x^2+b^2]^{3/2}}= \frac{q_2}{[x^2+c^2]^{3/2}} \to \frac{q_1}{q_2}=\frac{[x^2+b^2]^{3/2}}{[x^2+c^2]^{3/2}}
\end{equation}

\noindent As the ratio $\frac{q_1}{q_2}$ is constant, so must be the fraction $\frac{[x^2+b^2]^{3/2}}{[x^2+c^2]^{3/2}}$. Then, $b=c$ and therefore $q_1=q_2$.\\

\noindent From $b=c$, $q_1=q_2=q$, BC ${\bf D}_{2 \perp}={\bf D}_{1 \perp} +\sigma$, taking into account that there is no free surface charge ($\sigma=0$), and that ${\bf D}_{\perp}=\epsilon {\bf E}_{\perp}$, with ${\bf E}_{\perp}= -\frac{\partial V}{\partial z}$, we have:

\begin{equation}
{\bf E}_{1 \perp}(z>0) = \frac{Q}{\epsilon_1} \frac{z-a}{[x^2+(z-a)^2]^{3/2}}+ q \; \frac{z+b}{[x^2+(z+b)^2]^{3/2}}
\end{equation}
\begin{equation}
{\bf E}_{2 \perp}(z<0) = \frac{Q}{\epsilon_1} \frac{z-a}{[x^2+(z-a)^2]^{3/2}}+ q \; \frac{z-b}{[x^2+(z-b)^2]^{3/2}}
\end{equation}

\noindent Equating $\epsilon_1 {\bf E}_{1 \perp}=\epsilon_2 {\bf E}_{2 \perp}$ at $z=0$ we find:

\begin{equation}
- Q\; \frac{a}{[x^2+a^2]^{3/2}} + \epsilon_1 \; q \; \frac{b}{[x^2+b^2]^{3/2}}= - \frac{\epsilon_2}{\epsilon_1}\; Q \;\frac{a}{[x^2+a^2]^{3/2}}-\epsilon_2 \; q \;\frac{b}{[x^2+b^2]^{3/2}}
\end{equation}
\noindent Grouping factors:
\begin{equation}
Q\;(1-\frac{\epsilon_2}{\epsilon_1}) \frac{a}{[x^2+a^2]^{3/2}} = q (\epsilon_1+\epsilon_2) \;\frac{b}{[x^2+b^2]^{3/2}} \to \frac{\frac{1}{\epsilon_1}\; Q\;(\epsilon_1-\epsilon_2)}{q\; (\epsilon_1+\epsilon_2)} \frac{a}{b} = \frac{[x^2+a^2]^{3/2}}{[x^2+b^2]^{3/2}}.
\end{equation}
\noindent As the first term of the last equation is a constant, so must be the second, which implies $a=b$, a result that entails:

\begin{equation}
q=\frac{1}{\epsilon_1}\; \frac{(\epsilon_1-\epsilon_2)}{(\epsilon_1+\epsilon_2)} \;Q
\end{equation}

\noindent In summary, the electric potential generated by a charge $Q$ in a medium of dielectric constant $\epsilon_1$ located at a distance $a$ from the interface with another medium of dielectric constant $\epsilon_2$ is the same to that generated in a vacuum by an effective charge $Q^*=Q/\epsilon_1$ located at $Q$ site plus that generated by a image charge  $q=\frac{1}{\epsilon_1}\; \frac{(\epsilon_1-\epsilon_2)}{(\epsilon_1+\epsilon_2)} \;Q$ located symmetrically in the other medium at the same distance from the interface. In addition, $Q$ also generates in the other medium a potential like the one generated in a vacuum by an effective charge  $Q^*=Q/\epsilon_1$ plus an effective charge $q=\frac{1}{\epsilon_1}\; \frac{(\epsilon_1-\epsilon_2)}{(\epsilon_1+\epsilon_2)} \;Q$ both located at the $Q$ site.

\begin{center}
\begin{figurehere}
\resizebox{0.75\columnwidth}{!}{\includegraphics{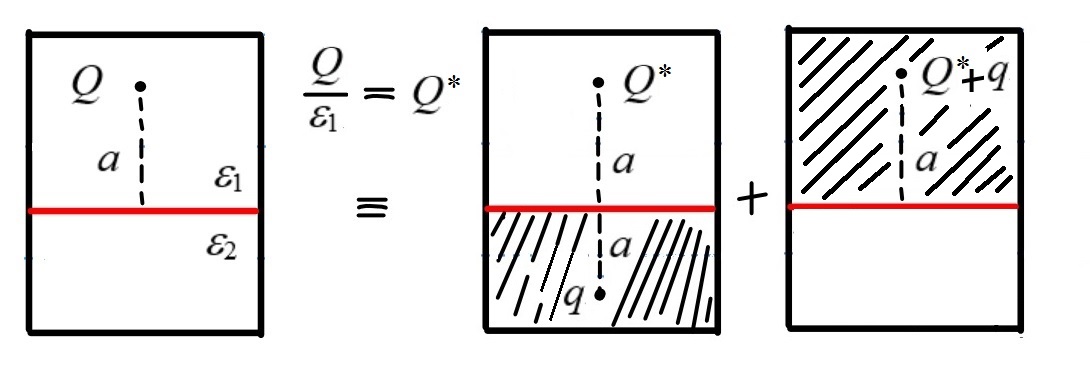}}
\end{figurehere}
\end{center}

\section{Appendix 3: On the electrodynamics atomic units}
There is some ambiguity in defining the atomic units of magnetic field and the own Maxwell equations. Within the {\it Lorentz force} convention, the Maxwell equations reads:
\begin{align}
{\boldsymbol \nabla}\cdot {\bf D}  &= 4 \pi \rho  & {\boldsymbol \nabla}\cdot  {\bf B} &=0\\
{\boldsymbol \nabla} \times {\bf E} &= -\frac{\partial {\bf B}}{\partial t} &  {\bf \nabla} \times {\bf H} &=\frac{\partial{\bf D}}{\partial t}+4 \pi\, {\bf J}
\end{align}

\noindent with the constitutive  relations ${\bf D}={\bf E} +  4 \pi {\bf P}$, ${\bf H}= c^2\, {\bf B}- 4 \pi {\bf M}$. In the last equation the value $c$ of the speed of light in a.u. is numerically equal to $\frac{1}{\alpha}$, with ${\alpha} = 1/137.036$ being the dimensionless fine-structure constant. These equations are complemented by the Lorentz force law: ${\bf F}= e \,({\bf E} + {\bf v}\times {\bf B})$.\\
  
\noindent Within the {\it Gaussian} convention, the Maxwell equations are:
\begin{align}
{\boldsymbol \nabla}\cdot {\bf D}  &= 4 \pi \rho  & {\boldsymbol \nabla}\cdot  {\bf B} &=0 \\
{\boldsymbol \nabla} \times {\bf E} &= -\alpha \frac{\partial {\bf B}}{\partial t} & {\bf \nabla} \times {\bf H} &= \alpha \frac{\partial{\bf D}}{\partial t}+ 4 \pi \alpha {\bf J}
\end{align}
\noindent with the constitutive  relations ${\bf D}={\bf E} +  4 \pi {\bf P}$, ${\bf H}= {\bf B}- 4 \pi {\bf M}$, which looks like those of the cgs-Gaussian system. In addition, we have the Lorentz force law: ${\bf F}= e \,({\bf E} + \frac{{\bf v}}{c}\times {\bf B})$.\\

\noindent Please note that the formal replacement $\alpha {\bf B} \to {\bf B}$, $\frac{{\bf H}}{\alpha} \to {\bf H}$, $\frac{{\bf M}}{\alpha} \to {\bf M}$ in the Gaussian convention equations retrieves those of the Lorentz  convention.\\

\noindent Care should we have in using either convention. For example, the atomic unit for magnetic field in the Lorentz force convention is $1 \,a.u._{LF}\approx2.35 \cdot 10^5 \, T$ while $1 \,a.u._{G}\approx 1.72 \cdot 10^7 \, G$. Since $1 \, T= 10^4 \, G$ we can check that $\frac{1 \, a.u._{LF}}{1 \, a.u._{G}}=\frac{1}{\alpha}$.\\

\noindent It would be instructive to have a look to the {\it units guide} provided by Andrea Dal Corso at the address:\\
\noindent https:$\backslash\backslash$people.sissa.it$\backslash\sim$dalcorso$\backslash$notes$\backslash$units.pdf $\;$ The text is motivated by the conversion factors implemented in the QUANTUM ESPRESSO (https:$\backslash\backslash$www.quantum-espresso.org/) but is of general interest.

\subsection{A bit more on electrodynamics units}
In order to dig on the different atomic units we start from two fundamental laws, the Coulomb force between electrical charges, $F= k_c \frac{q_1 \, q_2}{r^2}$ and the Biot-Savart force between infinitesimal $dl_1$,  $dl_2$ fragments of wires supporting current of intensity $i_1$ and $i_2$, respectively, $F= k_b \frac{i_1 \, i_2}{r^2} dl_1\, dl_2$. The constants $k_c$ and $k_b$ are related by the equation, 
\begin{equation}
\label{A30}
k_c/k_b=c^2,
\end{equation} 
\noindent with $c$ the speed of light.\\

\noindent We can assume this relationship as an experimental fact. All the same, it can be derived since, according to theory of relativity, magnetism is just electric interaction between moving charges (see e.g. chap. 5 in E.M. Purcell and J. Morin, {\it Electricity and Magnetism}, Cambridge University Press 2013).\footnote{ We can provide a simple, rather oversimplified, approach that can help to have a taste of this. To this end we consider $S_0$, a lab frame of reference attached to the positive cores of a wire supporting a current density $i_1$, carried by negative electrons. In this inertial frame we observe a neutral wire, i.e., the linear charge densities of cores $\rho_+$ and electrons $\rho_-$ are equal: $\rho_+ = \rho_-$. We employ the notation $L_0^+$ for the observed {\it proper} length between consecutive cores and  $L^-$ for the observed length between consecutive electrons  (remember: the {\it proper} length $L_0$ is the largest, so that the length $L$ observed from another inertial frame $S$, moving at the speed $v$ --where the system is not seen at rest--,  is related to $L_0$ by $L=L_0 \, \sqrt{1-v^2/c^2}$). We can write $\rho_+ = \frac{e}{L_0^+}$ and $\rho_- =- \frac{e}{L^-}$, with $e$ representing the absolute value of the electron charge. From $\rho_+ = -\rho_-$ we conclude,
\begin{equation}
\label{A31}
L_0^+ =L^-
\end{equation}
\noindent Next, we consider a second inertial frame $S'$ attached to electrons. Then, relative to $S_0$, it has a speed $v$. In this new frame the distance between consecutive {\it moving} cores is $L^+ = L_0 \sqrt{1-v^2/c^2}$  and that of electrons $L_0^-$. Interestingly, from this inertial frame we  see a net density charge in the wire:
\begin{equation}
\label{A32}
\rho'=\frac{e}{L^+}-\frac{e}{L_0^-} =\frac{e}{L_0^+}\frac{1}{\sqrt{1-v^2/c^2}}-\frac{e}{L^-} \, \sqrt{1-v^2/c^2}
\end{equation}
\noindent With eq. (\ref{A31}) we have:
\begin{equation}
\label{A33}
\rho'=\frac{e}{L_0^+}\frac{1}{\sqrt{1-v^2/c^2}} \, (1-1+\frac{v^2}{c^2}) = \frac{e}{L^+}\frac{v^2}{c^2}
\end{equation}

\noindent Since the density of cores observed in $S'$ is $\rho=\frac{e}{L^+}$, we can write:
\begin{equation}
\label{A34}
\rho'=\rho\frac{v^2}{c^2}.
\end{equation}

\noindent The cores are observed to move in $S'$ at speed $v$ then, the current intensity observed is $i = \frac{dq}{dt}=\rho \, \frac{dl}{dt} = \rho \, v$, so that $\rho'=i\, \frac{v}{c^2}$.\\

\noindent A  wire-length $dl$ with a non-zero density charge $\rho'$ yields an electrical field,
\begin{equation}
\label{A35}
E = \frac{k_c}{r^2} \rho' dl =  \frac{k_c}{r^2} \,i \, dl \, \frac{v}{c^2}.
\end{equation}

\noindent Let's write $i_1$ and $dl_1$ instead of  $i$ and $dl$ to indicate that they correspond to a given wire that we refer to as wire 1. \\

\noindent Consider next that $S'$ (attached to electrons of wire 1) sees a net charge $dq_2=\rho_2 dl_2$ moving along wire 2 at speed $v$. Then $S'$ sees a current intensity $i_2 = \rho_2 v$ in wire 2 and exert on it a force:
\begin{equation}
\label{A36}
F = E \, dq_2 = \frac{k_c}{r^2}  \, i_1 \, dl_1 \, \frac{v}{c^2} \, \rho_2 \, dl_2 = \frac{k_c}{c^2}  \, \frac{i_1 \, i_2}{r^2} \, dl_1 \, dl_2,
\end{equation}

\noindent that, if compared to the Biot-Savart force, $F= k_b \frac{i_1 \, i_2}{r^2} dl_1\, dl_2$, lead us to say: $\frac{k_c}{c^2} = k_b$.
}
\noindent Once we have settled eq. (\ref{A30}), there are several options to define these constants, yielding the different unit systems. The cgs Gaussian unit system assumes $k_c=1$ and, according to  (\ref{A30}), $k_b=1/c^2$. The same does the atomic unit system in the so-called Gaussian convention. However, while $k_c$ has the same numerical value in both unit systems, taking into account that $c = 2.89\cdot 10^{10} \; cm/s$ and $c = 137.036 \; a.u.$, $k_b$ has a different numerical value in either system.\\

\noindent On the other hand, the international SI or MKS unit system assumes $k_c=\frac{1}{4 \pi \epsilon_0}$ and $k_b= \frac{\mu_0}{4 \pi}$ with  $\epsilon_0=8.854\cdot 10^{-12} \; C^2/N\cdot m^2$, $\mu_0= 4 \pi \cdot 10^{-7} \; N/A^2$.  Again, $\frac{k_c}{k_b} = \frac{1}{\epsilon_0 \mu_0} = 8.988\cdot 10^{16}\; m^2/s^2 = c^2$.\\

\noindent Another challenge is the transformation of electromagnetic equations between both unit systems. To this end we first invoke the Coulomb equation:
$$
\frac{q^G_1 \, q^G_2}{r^2} \hspace{1cm} vs. \hspace{1cm} \frac{1}{4 \pi \epsilon_0} \frac{q^{SI}_1 \, q^{SI}_2}{r^2} = \frac{\frac{q^{SI}_1}{\sqrt{4 \pi \epsilon_0}} \frac{q^{SI}_2}{\sqrt{4 \pi \epsilon_0}}}{r^2}
$$ 

\noindent where $r$ must be expressed either, in $cm$ (Gaussian formula) or in $m$ (SI formula). Then, we see that 
\begin{equation}
\label{A37}
q^{SI}=\sqrt{4 \pi \epsilon_0} \, q^G. 
\end{equation}

\noindent We consider next the electric field:
\begin{equation}
\label{A38}
E^{SI}=\frac{1}{4 \pi \epsilon_0} \, \frac{q^{SI}}{r^2} = \frac{1}{\sqrt{4 \pi \epsilon_0}} \, \frac{q^G}{r^2} = \frac{1}{\sqrt{4 \pi \epsilon_0}} \, E^G
\end{equation}

\noindent In a similar way, we derive the potential:
\begin{equation}
\label{A39}
V^{SI} = \frac{V^G}{\sqrt{4 \pi \epsilon_0}}
\end{equation}

\noindent In dielectric media, the Coulomb law turns into $ F =  \frac{1}{4 \pi \epsilon_0} \, \frac{1}{\epsilon'} \frac{q^{SI}_1 \, q^{SI}_2}{r^2} = \frac{1}{\epsilon'} \frac{q^G_1 \, q^G_2}{r^2}$. Then,

\begin{equation}
\label{A40}
\epsilon^{SI} = \epsilon_0 \epsilon^G =  \epsilon_0 \epsilon'   \hspace{1cm} i.e.   \hspace{1cm} \epsilon^G = \epsilon'
\end{equation}

\noindent Finally,

\begin{equation}
\label{A41}
 D^{SI} = \epsilon^{SI} E^{SI} =  \epsilon_0  \epsilon^G \, \frac{1}{\sqrt{4 \pi \epsilon_0}}\, E^G = \sqrt{\frac{ \epsilon_0}{4 \pi}} \; D^G.
\end{equation}

\noindent In the magnetic equations  we have $\frac{\mu_0}{4 \pi}$  instead of $\frac{1}{4 \pi \epsilon_0}$ and poles instead of charges. Then, proceeding in a similar way, we find out:
  
\begin{eqnarray}
\label{A42}
p^{SI}= \sqrt{\frac{4 \pi}{\mu_0}} \, p^G \\
B^{SI} = \sqrt{\frac{\mu_0}{4 \pi}} \, B^G \\
U^{SI} = \sqrt{\frac{\mu_0}{4 \pi}} \, U^G \\
\mu^{SI}= \mu_0 \mu^G \\
H^{SI} = \frac{1}{\sqrt{4 \pi \mu_0}} \, H^G
\end{eqnarray}

\noindent A complete table with Gaussian-SI equivalences and constant numerical values is reported by wikipedia (https:$\backslash\backslash$en.wikipedia.org$\backslash$wiki$\backslash {\rm Gaussian\_units}$).\\

\noindent With these equation we can translate eqs. in section 3.2 from Gaussian (or atomic unit in Gaussian convention) to SI units. I enclose, next, some examples.
\begin{enumerate}
\item Equation (\ref{eqimag3}):\\
$U^G(x,z>0) = \frac{p^G_2}{[x^2+(z+b)^2]^{1/2}}$ $  \to  $  $U^{SI} = \sqrt{\frac{\mu_0}{4 \pi}} U^G = \sqrt{\frac{\mu_0}{4 \pi}}\frac{\sqrt{\frac{\mu_0}{4 \pi}}\,p^{SI}_2}{[x^2+(z+b)^2]^{1/2}} =\frac{\mu_0}{4 \pi} \frac{p^{SI}_2}{[x^2+(z+b)^2]^{1/2}}$
\item Boundary condition  ${\bf D}^G_{1 \perp} = {\bf D}^G_{2 \perp} -\alpha B^G_z$ yields  ${\bf D}^{SI}_{1 \perp} = {\bf D}^{SI}_{2 \perp} - \sqrt{\frac{\epsilon_0}{\mu_0}} \alpha B^{SI}_z$ \\
      i.e. $\epsilon^{SI}_1 {\bf E}^{SI}_{1  \perp}=\epsilon^{SI}_2 {\bf E}^{SI}_{2 \perp}- \sqrt{\frac{\epsilon_0}{\mu_0}} \alpha B^{SI}_z$. 
\item Boundary condition  ${\bf H}^G_{1 \parallel} = {\bf H}^G_{2 \parallel} +\alpha {\bf E}^G_{2 \parallel}$ yields ${\bf H}^{SI}_{1 \parallel} = {\bf H}^{SI}_{2 \parallel} +\sqrt{\frac{\epsilon_0}{\mu_0}} \alpha {\bf E}^{SI}_{2 \parallel}$ \\
      i.e. $\frac{1}{\mu^{SI}_1} {\bf B}^{SI}_{1 \parallel} =\frac{1}{\mu^{SI}_2} {\bf B}^{SI}_{2 \parallel} +\sqrt{\frac{\epsilon_0}{\mu_0}} \alpha {\bf E}^{SI}_{2 \parallel}$
\item Equation (\ref{eq16}):  $p^G_1= \frac{\alpha}{\frac{1}{\mu^G_1}+\frac{1}{\mu^G_2}} \;(\frac{Q^G}{\epsilon^G_1}+q^G)$ with $ p^{SI}_1=\sqrt{\frac{4 \pi}{\mu_0}} \,p^G_1$ yields,\\
     $p^{SI}_1= \sqrt{\frac{4 \pi}{\mu_0}} \frac{\alpha}{\frac{\mu_0}{\mu^{SI}_1}+\frac{\mu_0}{\mu^{SI}_2}}\; \left(\frac{1}{\sqrt{4 \pi \epsilon_0}} \frac{Q^{SI}}{\epsilon'_1}+ \frac{1}{\sqrt{4 \pi \epsilon_0}} q^{SI} \right) =
	   \sqrt{\frac{ \epsilon_0}{\mu_0}} \frac{\alpha/\mu_0}{\frac{1}{\mu^{SI}_1}+\frac{1}{\mu^{SI}_2}} \left(  \frac{Q^{SI}}{\epsilon^{SI}_1}+ \frac{q^{SI}}{\epsilon_0} \right)$
\item Equation (\ref{eq17}): $q^G = \frac{1}{\sqrt{4 \pi \epsilon_0}} q^{SI}=\frac{Q^G}{\epsilon^G_1} \frac{(\frac{1}{\mu^G_1}+\frac{1}{\mu^G_2}) (\epsilon^G_1-\epsilon^G_2)-\alpha^2}{(\frac{1}{\mu^G_1}+\frac{1}{\mu^G_2}) (\epsilon^G_1+\epsilon^G_2)+\alpha^2}$ yields,\\
	 $q^{SI} = \epsilon_0 \; \frac{Q^{SI}}{\epsilon^{SI}_1} \frac{(\frac{1}{\mu^{SI}_1}+\frac{1}{\mu^{SI}_2}) (\epsilon^{SI}_1-\epsilon^{SI}_2)-\frac{\epsilon_0}{\mu_0} \,\alpha^2}{(\frac{1}{\mu^{SI}_1}+\frac{1}{\mu^{SI}_2}) (\epsilon^{SI}_1+\epsilon^{SI}_2)+ \frac{\epsilon_0}{\mu_0} \,\alpha^2}$
\item Finally, Equation (\ref{eq18}): $p^G_1= \alpha \, \frac{2 Q^G}{(\frac{1}{\mu^G_1}+\frac{1}{\mu^G_2})(\epsilon^G_1+\epsilon^G_2)+\alpha^2}$ yields,\\
	$p^{SI}_1= \frac{1}{\mu_0} \left( \sqrt{\frac{\epsilon_0}{\mu_0}} \, \alpha \right) \, \frac{2 Q^{SI}}{(\frac{1}{\mu^{SI}_1}+\frac{1}{\mu^{SI}_2})(\epsilon^{SI}_1+\epsilon^{SI}_2)+  \frac{\epsilon_0}{\mu_0} \, \alpha^2}$
\end{enumerate}

\noindent As far as the dimensionless  fine-structure constant $\alpha$, it has the same numerically value in all unit systems ($\alpha=1/137.036$), but presents different formulas in different systems. Thus, $\alpha^{SI}=\frac{e^2}{4 \pi \epsilon_0 \hbar c} = 7.297 \cdot 10^{-3}$. Since  $\epsilon^G_0 = \frac{1}{4 \pi}$, then  $\alpha^G=\frac{e^2}{\hbar c} = 7.297 \cdot 10^{-3}$. Finally, from $e=\hbar=1 \,a.u.$, $\alpha^{a.u.}=\frac{1}{c} = 7.297 \cdot 10^{-3}$.

\begin{thebibliography}{99}
\bibitem{NotaLagrangiana} The electromagnetic energy density $W = \frac{1}{2 \mu} B^2+\frac{1}{2} \epsilon E^2$ can formally be identified with a momentum $B$, mass $\mu$, coordinate $E$ and force constant $\epsilon$ harmonic oscil$\cdot$lator, so that energy $W$ is a sum of kinetic $T$ and potential $V$ energies. By convention, Lagrangian is defined as ${\cal L}=T-V$, but can also be defined with opposite sign  ${\cal L}=V-T=\frac{1}{2} \epsilon E^2 - \frac{1}{2} \frac{1}{\mu} B^2$ for both definitions lead to the same Euler-Lagrange equation $\frac{d}{dt}\left(\frac{\partial {\cal L}}{\partial \dot{q} }\right)-\left(\frac{\partial {\cal L}}{\partial q }\right)=0$.
\bibitem{nota2} Since for image charges calculation only equations (\ref{eq1b}) i (\ref{eq4b})  are employed, Visinelli proposal\cite{Visinelli} does not affect this kind of calculation.
\end{thebibliography}
\end{document}